\documentclass[aps,prb,superscriptaddress,twocolumn,floatfix,showpacs,a4paper]{revtex4-1}

\usepackage{amsfonts,amsmath,amssymb}
\usepackage{textcomp}
\usepackage{bm}
\usepackage{upgreek}
\usepackage{graphicx}
\usepackage[latin1]{}
\newcommand{\ie}{\textit{i.e.}}

\newcommand{\etc}{\textit{etc.}}
\newcommand{\etal}{\textit{et~al.}}

\renewcommand{\vec}[1]{\bm{#1}}
\newcommand{\ee}{\mathrm{e}}
\newcommand{\ii}{\mathrm{i}}

\newcommand{\nablabf}{\boldsymbol{\nabla}}

\newcommand{\rot}{\nablabf\times}

\newcommand{\fracsmall}[2]{\mbox{$\frac{#1}{#2}$}}

\newcommand{\AAA}{\vec{A}}

\newcommand{\aaa}{\vec{a}}

\newcommand{\GGn}{\vec{G}}

\newcommand{\ggn}{\vec{g}}

\newcommand{\HHH}{\vec{H}}

\newcommand{\kkk}{\vec{k}}

\newcommand{\rrr}{\vec{r}}

\newcommand{\uuu}{\vec{u}}

\newcommand{\VVV}{\vec{V}}

\newcommand{\calT}{\mathcal{T}}

\newcommand{\eps}{\epsilon}

\newcommand{\beq}[1]{\begin{equation} \eqlab{#1}}
\newcommand{\eeq}{\end{equation}}
\newcommand{\bsub}{\begin{subequations}}
\newcommand{\esub}{\end{subequations}}
\def\bal#1\eal{\begin{align}#1\end{align}}
\def\bsubal#1\esubal{\bsub \begin{align}#1\end{align} \esub}

\newcommand{\eqlab}[1]{\label{eq:#1}}
\renewcommand{\eqref}[1]{Eq.~(\ref{eq:#1})}
\newcommand{\eqsref}[2]{Eqs.~(\ref{eq:#1}) and~(\ref{eq:#2})}

\newcommand{\figref}[1]{Fig.~\ref{fig:#1}}

\newcommand{\appref}[1]{Appendix~\ref{sec:#1}}

\newcommand{\secref}[1]{Section~\ref{sec:#1}}

\newcommand{\seclab}[1]{\label{sec:#1}}

\begin{document}

\title{Patched Green's function techniques for two dimensional systems: Electronic behaviour of bubbles and perforations in graphene}

\author{Mikkel Settnes}\email[]{mikse@nanotech.dtu.dk}
\author{Stephen R. Power}
\author{Jun Lin}
\author{Dirch H. Petersen}
\author{Antti-Pekka Jauho}
\affiliation{Center for Nanostructured Graphene (CNG), DTU Nanotech, Technical University of Denmark, DK-2800 Kongens Lyngby, Denmark}

\date{\today}

\begin{abstract}
We present a numerically efficient technique to evaluate the Green's function for extended two dimensional systems without relying on periodic boundary conditions.  
Different regions of interest, or `patches', are connected using self energy terms which encode the information of the extended parts of the system.
The calculation scheme uses a combination of analytic expressions for the Green's function of infinite pristine systems and an adaptive recursive Green's function technique for the patches.
The method allows for an efficient calculation of both local electronic and transport properties, as well as the inclusion of multiple probes in arbitrary geometries embedded in extended samples.
We apply the Patched Green's function method to evaluate the local densities of states and transmission properties of graphene systems with two kinds of deviations from the pristine structure: bubbles and perforations with characteristic dimensions of the order of 10-25 nm, \textit{i.e.} including hundreds of thousands of atoms.
The  strain field induced by a bubble is treated beyond an effective Dirac model, and
we demonstrate the existence of both Friedel-type oscillations arising from the edges of the bubble, as well as pseudo-Landau levels related to the pseudomagnetic field induced by the nonuniform strain.
Secondly, we compute the transport properties of a large perforation with atomic positions extracted from a TEM image, and show that current vortices may form near the zigzag segments of the perforation.
\end{abstract}

\maketitle

\section{Introduction}
Following the isolation of graphene a general class of two dimensional materials with widely diverse and unique electrical, mechanical and optical properties has been realized. \cite{Geim2013,Fiori2014}
Two dimensional materials are almost entirely surface and are therefore very susceptible to external influences like direct patterning \cite{Bai2010}, adsorbate atoms \cite{Schedin2007}, strain \cite{Qi2013}, \etc \;
This variety of ways to alter and control the material properties opens a huge range of engineering possibilities.\cite{Novoselov2012}
In this context, it becomes important to investigate large scale disorder or patterning in relation to the electronic properties of graphene and related two dimensional materials.
From a theoretical perspective several methods are available. \cite{CastroNeto2009}
Typically, the electronic structure of the system is described with a tight-binding type Hamiltonian and a popular approach is then to construct the entire system in a piece-wise manner using recursive Green's functions (RGFs).\cite{Lewenkopf2013} In this way, we can extract the necessary terms for calculating physical quantities of interest.
The RGF method is best-suited for systems which are either finite or periodic in one dimension.
It is frequently used for modeling transport, where self energies calculated using recursive techniques are used to attach semi-infinite pristine leads to either side of a finite device region. \cite{DattaBook}
Alternatively, an efficient approach to large disordered systems is the real space Kubo-Greenwood approach. \cite{RocheBook}
However, this method cannot include open boundary conditions and can only obtain average system quantities, as opposed to local electronic and transport properties.

In the most common formulation, the RGF method treats (quasi) one dimensional systems with only two leads.
Although variants of the method can be used for arbitrary geometries and multiple leads, \cite{Kazymyrenko2008,Wimmer2009} the method remains limited to finite-width or periodic systems.
Consequently, it cannot describe local and non-periodic perturbations, or point-like probes similar to those considered experimentally. \cite{Baringhaus2014,Sutter2008}
An extension of recursive techniques, to allow efficient treatment of local properties in systems without periodicity or finite sizes, would allow for easier theoretical investigation of systems which are computationally very expensive, or completely out of reach, using existing methods.

%The electronic properties of graphene and related two dimensional materials is of particular interest and can be investigated through a variety of theoretical methods. \cite{CastroNeto2009} In particular Green's function (GF) methods combined with tight binding type Hamiltonians is a popular approach. In many applications, we are interested in conductance, where geometry plays a crucial role. A standard way to treat this kind of transport is the recursive Green's function approach, \cite{Lewenkopf2013} which has proven useful for a wide range of mesoscopic systems. \cite{DattaBook} However, in its original formulation it is intrinsically (quasi) one dimensional with only two leads. Although variants of the method can be used for arbitrary geometries and multiple leads, \cite{Kazymyrenko2008,Wimmer2009} the method remains limited to finite systems or systems with a period in the transverse direction. This is problematic if we want to treat the influence of individual perturbations within a larger sample like local deformations \cite{Qi2013,Levy2010} or if we consider local probes \cite{Baringhaus2014,Sutter2008} essentially coupling to points on a larger sample.

\begin{figure*}[htb]
\includegraphics[width=0.9\columnwidth]{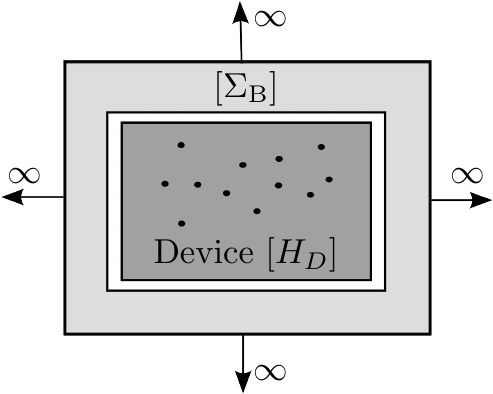}\hspace{2cm}
\includegraphics[width=0.9\columnwidth]{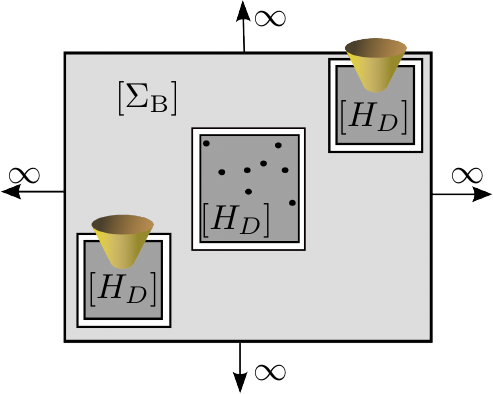}
\caption[]{The left panel shows a schematic of a computational setup containing a finite device `patch' , described by $\HHH_D$, embedded within an extended system described by the self energy $\mathbf{\Sigma}_B$. The right panel shows a computational setup containing several device `patches' of interest.} \label{fig:setup}
\end{figure*}

In this paper, we develop a Green's function (GF) method which is able to efficiently treat large and finite sized 'patches' embedded in an extended system, as shown in \figref{setup}.
The method combines an analytical formulation of the Green's functions describing a pristine system \cite{Settnes2014a,Power2011} with an adaptive recursive Green's function method
to described the patches.
It allows for calculation of both local electronic and transport properties and for the inclusion of multiple leads and arbitrary geometries embedded within an extended sample.

This patched Green's function method exploits an efficient calculation of the GF for an infinite pristine system using complex contour techniques.
Using this GF, an open boundary self energy term can be included in the device Hamiltonian to describe its connection to an extended sample.
The device region itself, containing nanostructures and/or leads, is then treated with an adaptive recursive method.
We demonstrate the formulation using graphene, but it is generally applicable to all (quasi) 2D structures where the Green's function for the infinite pristine system can be determined. Consequently, the patched Green's function method is a versatile tool for efficient investigation of non-periodic nanostructures in extended two dimensional systems.

The rest of the paper is outlined as follows: the general formalism is developed in \secref{SquareSelfEnergy} by calculating the open boundary self energy from the pristine GF. In \secref{Gint} we use graphene as an example to show the calculation of the pristine GF, while \secref{recursive} discusses the adaptive recursive method used to treat the device when including the boundary self energy. In \secref{bubble} we use the developed method to study the local density of states of a graphene sample under the influence of a local strain field. As a result, we can compare local density of state (LDOS) maps with pseudomagnetic field distributions. In this way, we show the existence of Friedel-type oscillations along with pseudomagnetic field effects in the LDOS. Finally, in \secref{antidot}, we use the patched Green's function technique to demonstrate the existence of vortex like current patterns in the presence of a perforation within an extended graphene sheet.

\begin{figure*}[htb]
\includegraphics[width=0.9\textwidth]{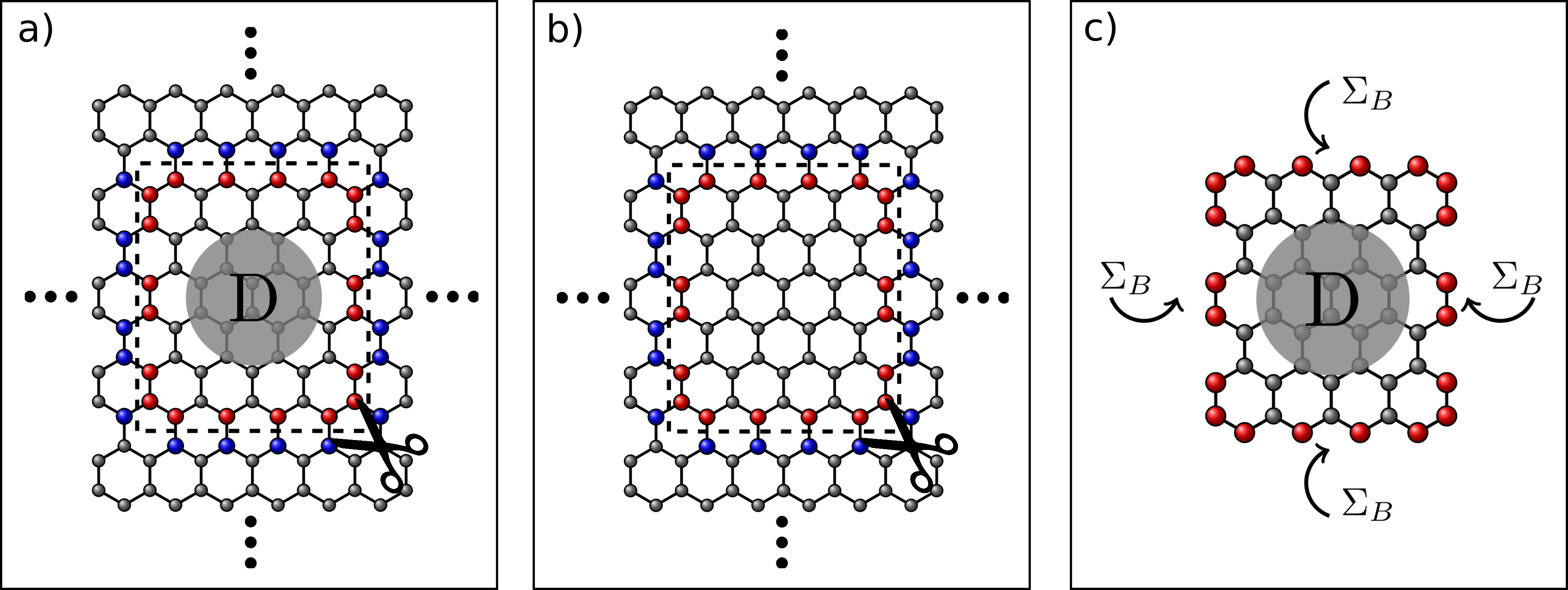}
\caption[]{ a) Shows the desired device region, indicated by the dashed square, embedded within an extended system. Red symbols are the edge of the device and blue symbols indicate sites in the surrounding sheet that couples to the device. We obtain the disconnected system discussed in the text by removing the couplings that cross the dashed line. b) Shows the corresponding pristine system. Again the disconnected system is obtained by removing couplings along the dashed line. c) Illustrates how the effect of the extended sheet on the device region is taken into account by the self energy, see \eqref{sigmaB}.} \label{fig:SSE_sketch}
\end{figure*}

\section{Method}\seclab{method}
We consider the computational setup schematically shown in the left panel of \figref{setup}, where a device region is embedded within an extended two dimensional system.
This setup ensures that we are not including edge effects due to the finite-size of the simulation domain. \cite{Sajjad2013}
The device region is described by a Hamiltonian, $\HHH$, which may include disorder, deformations, mean field terms or leads \etc \;
This device region is embedded into an extended system by applying a self energy term, $\Sigma_B$. 
To consider the setup in \figref{setup}, we need two things: first, we need to construct $\Sigma_B$ to describe the extended part of the system and secondly, we need an efficient way to describe the device region while taking $\Sigma_B$ into account.
Furthermore, the treatment of the device should be able to consider arbitrary geometries, including mutually disconnected patches within the extended system, as shown in the right panel of \figref{setup}.

We describe the method in three steps:
\begin{itemize}
	\item[A:] Derivation of the boundary self energy term, $\Sigma_B$, in terms of the %connected and disconnected 
pristine lattice GFs. 
	\item[B:] Calculation of the real-space GF needed in the self energy calculation. We use graphene as an example.
	\item[C:] Implementation of an adaptive RGF method to build the device region(s) efficiently while including the self energy term(s) $\Sigma_B$. 
\end{itemize}

%The self energy term, $\Sigma_B$ is derived in \secref{SquareSelfEnergy} and require the real space GF for the pristine and infinite system. Consequently, we dedicate \secref{Gint} to calculate this GF using graphene as an example.
%To treat the device and determine specific GF terms between points of interest, we outline an adaptive RGF method in \secref{recursive} taking into account an arbitrary choice of interest points as well as the non-hermitian $\Sigma_B$.

\subsection{Boundary self energy}\seclab{SquareSelfEnergy}
To construct the boundary self energy describing the extended region in \figref{setup}, we consider the simple graphene example in \figref{SSE_sketch}a. Here a central device region, indicated by the dashed square, is embedded into an extended sheet.
In this example both the extended area and the device region are assumed to be graphene-based, but the following arguments are general to any two dimensional material.
We consider a division of the system into two parts: sites in the device (D) or sites in the extended sheet region.
Furthermore, we subdivide the extended sheet into boundary sites (B) which are indicated by blue in \figref{SSE_sketch} and have a non-zero Hamiltonian element coupling them to the device region, or `sheet' sites which do not couple to the device region.
Within a nearest-neighbour tight-binding Hamiltonian, the boundary sites in \figref{SSE_sketch}a are shown by blue symbols and have non-zero couplings to the device sites indicated by red symbols.
We can now write the Hamiltonian for the entire system, in block matrix form, as
\bal
\HHH = \left(\begin{array}{ccc}
	 \HHH_{D,D}   &   \VVV_{D,B}  & 0 \vspace{0.1cm}\\
	\VVV_{B,D}   &  \HHH_{B,B} & \VVV_{B,\mathrm{sheet}}  \\
	0   &  \VVV_{\mathrm{sheet},B}   & \HHH_{\mathrm{sheet}}
\end{array}\right)\eqlab{H},
\eal
where the light shaded part of \eqref{H} represent an infinite Hamiltonian. The connections between device and sheet, (i.e. between the red and blue symbol sites) are contained in the off-diagonal blocks $\VVV_{D,B}$ and $\VVV_{B,D}$.

We aim to replace the infinite Hamiltonian $\HHH$  with a finite effective Hamiltonian, $\HHH_{\rm{eff}} = \HHH_{D,D} + \mathbf{\Sigma}_B$, which takes into account the extended sheet using a self energy term $\mathbf{\Sigma}_B$.
To do this, we consider the \emph{connected} system in panel a) of \figref{SSE_sketch}, and a disconnected system formed by removing the Hamiltonian elements $\VVV_{D,B}$ and $\VVV_{B,D}$, corresponding to removing couplings crossing the dashed line in \figref{SSE_sketch}a.
The GFs of the connected ($\GGn^{(\mathrm{con})}$) and disconnected ($\GGn^{(\mathrm{dis})}$) systems can be related via the Dyson equation, and in particular we can write the GF of the connected device region as
%
%The GFs of the disconnected system and the connected system on \figref{SSE_sketch} (Step 1) is related through a Dyson equation using the coupling $\VVV_{D,B}$ as perturbation,
\bal \eqlab{eqn_disc}
\GGn^{(\mathrm{con})}_{D,D} &= \GGn^{(\mathrm{dis})}_{D,D} +\GGn^{(\mathrm{dis})}_{D,D} \VVV_{D,B}\GGn^{(\mathrm{con})}_{B,D}. %,\\
%\GGn^{(\mathrm{con})}_{B,D} &= \GGn^{(\mathrm{dis})}_{B,B} \VVV_{B,D}\GGn^{(\mathrm{con})}_{D,D}. \\%\GGn^{\mathrm{con}}_{D,D} &= \GGn^{\mathrm{dis}}_{D,D} +\GGn^{\mathrm{dis}}_{D,D} \VVV_{D,B}\GGn^{\mathrm{dis}}_{B,B} \VVV_{B,D}\GGn^{\mathrm{con}}_{D,D} 
%\GGn^{\mathrm{dis}}_{D,D}&=(1-\GGn^{\mathrm{dis}}_{D,D} \VVV_{D,B}\GGn^{\mathrm{dis}}_{B,B} \VVV_{B,D})\GGn^{\mathrm{con}}_{D,D}    \\
%\GGn^{\mathrm{con}}_{D,D} &= \frac{\GGn^{\mathrm{dis}}_{D,D}}{1-\GGn^{\mathrm{dis}}_{D,D} \VVV_{D,B}\GGn^{\mathrm{dis}}_{B,B} \VVV_{B,D}}  \\
%\GGn^{\mathrm{con}}_{D,D} &= \frac{1}{(\GGn^{\mathrm{dis}}){}^{-1}- \VVV_{D,B}\GGn^{\mathrm{dis}}_{B,B} \VVV_{B,D}}  \\
%\GGn^{\mathrm{con}}_{D,D} &= \frac{1}{(E-\HHH_{D,D})- \VVV_{D,B}\GGn^{\mathrm{dis}}_{B,B} \VVV_{B,D}}  \\
\eal
Applying the Dyson equation again to obtain $\GGn^{\mathrm{(con)}}_{B,D}$ and inserting this into \eqref{eqn_disc} allows us to simplify,
\bal
\GGn^{(\mathrm{con})}_{D,D} &= \big(E\mathbf{1}-\HHH_{D,D}- \mathbf{\Sigma}_B \big)^{-1}  ,
 \eqlab{Gdd}
 \eal
where the self energy term is given by
 \bal
\mathbf{\Sigma}_B &=  \VVV_{D,B}\GGn^{(\mathrm{dis})}_{B,B} \VVV_{B,D}.\eqlab{sigmaB}
\eal

We note that the self energy in \eqref{sigmaB} is independent of the considered device and depends only on GF matrix elements connecting sites in the pristine surrounding `frame' that remains when the device is removed from the full system.
We take advantage of this to temporarily replace the device with a corresponding pristine region of the same size, as shown in panel b) of \figref{SSE_sketch}.
The self-energy required to incorporate the finite pristine region into an infinite, pristine sheet is the same self energy, $\mathbf{\Sigma}_B$, that is required in \eqref{Gdd}.
We can therefore write the required GF matrix, $\GGn^{(\mathrm{dis})}_{B,B}$, in terms of the GF of the infinite pristine sheet, $\GGn^{(\mathrm{0})}$. These are related using the Dyson equation with a perturbation $-\VVV_{D,B}$,

\bal
% \GGn^{(\mathrm{dis})}_{B,B} &= \GGn^{(\mathrm{con})}_{B,B} + \GGn^{(\mathrm{con})}_{B,A} (-\VVV_{A,B}) \GGn^{(\mathrm{dis})}_{B,B} ,\\
 \GGn^{(\mathrm{dis})}_{B,B} &=\big(\mathbf{1} + \GGn^{(\mathrm{0})}_{B,D} \VVV_{D,B}\big)^{-1} \GGn^{(\mathrm{0})}_{B,B}. \eqlab{solDysonGbb}
\eal
The advantage of writing the self-energy in terms of the pristine sheet GFs, $\GGn^{(0)}_{B,B}$ and $\GGn^{(0)}_{B,D}$, becomes clear in the next section, where we demonstrate an efficient method to calculate these two terms.
It is worth noting that $\GGn^{(0)}_{B,D}$ only needs to be calculated for the sites in D which connect to sites in B.
These sites are indicated by red in \figref{SSE_sketch} and are where the self-energy terms need to be added, as shown in panel c). In this way, the computations only involve matrices corresponding to the edge of the device and not the size of the full device region as straight forward inversion would require.

The calculation scheme can be summarized as follows:

\begin{itemize}
\item[ 1:] Calculate $\GGn^{(\mathrm{0})}_{B,B}$ and $\GGn^{(\mathrm{0})}_{B,D}$ using the methods outlined in \secref{Gint}.%\\ (GF for blue sites and connection between blue and red sites, see \figref{SSE_sketch}).
\item[ 2:] Calculate  $\mathbf{\Sigma}_B$ from  \eqref{sigmaB} and \eqref{solDysonGbb}.
\item[ 3:] The finite GF for the device region, $\GGn^{(\mathrm{con})}_{D,D}$, is given by \eqref{Gdd} and can be treated using an adaptive RGF method, see \secref{recursive}.
\end{itemize}

We note that this approach does not require a specific geometric shape of the device, nor does the device region need to be contiguous.
We can treat different non-connected patches in an extended system, as shown in the right panel of \figref{setup}, by extending the set D to include sites inside each patch and similarly expanding B to include sites at the boundary of each patch.
The method presented in this section is applicable to any system where the connected, pristine GFs are easily obtainable as demonstrated in the next section using a tight-binding description of graphene as an example.
%In practice this requires that $\GGn^{(\mathrm{con})}_{B,B}$ and $\GGn^{(\mathrm{con})}_{B,D}$ can be obtained in an analogous manner to that shown in the next section using a tight-binding description of graphene as an example.

\subsection{Real space graphene Green's function}\seclab{Gint}
We now turn to the calculation of the real space GF of the pristine system, which is needed to calculate the self energy, $\Sigma_B$, in \eqref{sigmaB} and \eqref{solDysonGbb}.
The approach required to calculate this quantity is demonstrated below for the case of a graphene sheet described with a nearest-neighbor tight-binding Hamiltonian, but is easily generalized for other cases.

This Hamiltonian is given by
\bal
\HHH = \sum_{<i,j>} t\hat{c}_i^\dagger \hat{c}_j \eqlab{H0},
\eal
where the sum $<i,j>$ runs over all nearest neighbour pairs and the carbon-carbon hopping integral is $t\approx -2.7$ eV.
The graphene hexagonal lattice can be split into two triangular sublattices, which we denote A and B, and neighbouring sites reside on opposite sublattices to each other.
Using Bloch functions, the Hamiltonian can be rewritten in reciprocal space as \cite{CastroNeto2009}
\bal
\HHH_{\kkk} = t \begin{pmatrix}
0 & f(\kkk)\\
f^*(\kkk) & 0
\end{pmatrix},
\eal
where the matrix form arises from sublattice indexing within a 2 atom unit cell and we have used the definition $f(\kkk) = 1 + \ee^{\ii \kkk \cdot\aaa_1} + \ee^{\ii \kkk \cdot\aaa_2}$, with the lattice vectors $\aaa_1 = a_0 ( \sqrt{3},3)/2$ and $\aaa_2 = a_0 (-\sqrt{3},3)/2$ and $a_0$ the carbon-carbon distance. With this definition of the unit vectors we have the armchair direction along the y-axis (and zigzag along the x-axis).

The eigenenergies and eigenstates of the system are easily obtained from this form of the Hamiltonian, and transforming back to real space allows us to write the desired Green's function between sites $i$ and $j$ as \cite{Bena2009,Power2011}
\bal
G^0_{ij}(z) = \frac{1}{\Omega_{BZ}}\int\mathrm{d}^2\kkk \frac{N_{ij}(z,\kkk)\ee^{\ii \kkk\cdot (\rrr_j-\rrr_i)}}{z^2 - t^2|f(\kkk)|^2}, %\begin{pmatrix}
%z & tf(\kkk) \\
%t f^*(\kkk) & z
%\end{pmatrix} ,
\eqlab{g0}
\eal
where $z = E +\ii 0^+$ is the energy, $\Omega_{BZ}$ is the area of the first Brillouin zone. The position of the unit cell containing site $i$ is denoted by $\rrr_{i} = m_{i}\aaa_1+n_{i} \aaa_2$ with $m_i$ and $n_i$ being integers. Finally we use the definition $N_{ij}(z,\kkk) = z$, when $i$ and $j$ are on the same sublattice and $N_{ij}(z,\kkk) = tf(\kkk)$ if $i$ is on the A sublattice and $j$ is on the B sublattice and $N_{ij}(z,\kkk) = tf^*(\kkk)$ when $i$ is on B and $j$ on A.

\begin{figure*}[htb]
\includegraphics[width=0.85\textwidth]{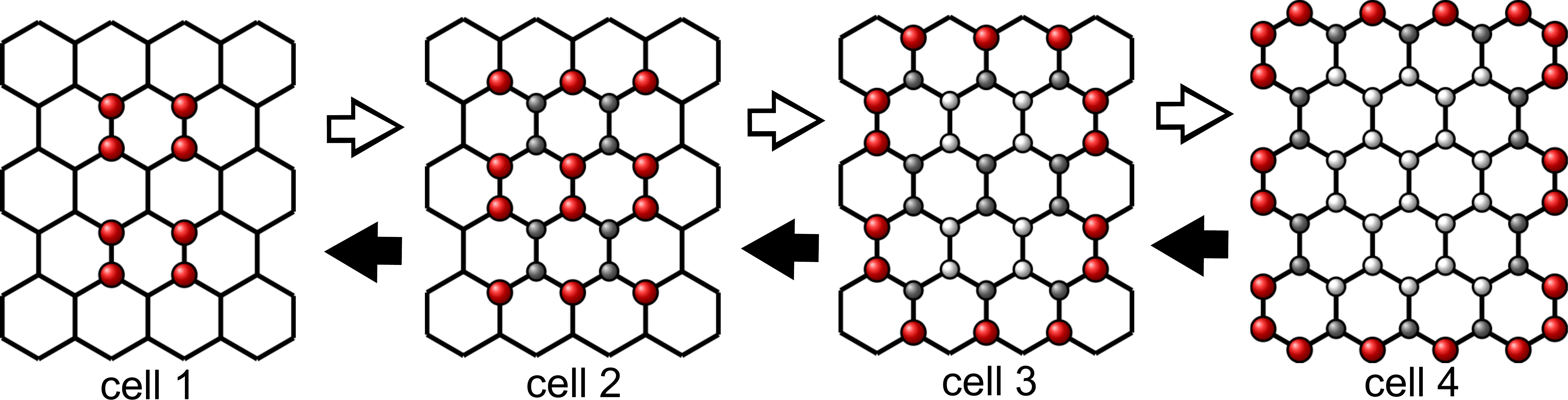}
\caption[]{ The partitioning of a small graphene sample where all sites of interest are located in cell 1. Cell 2 contains all the sites coupling to cell 1 but which are not themselves part of cell 1. Likewise cell 3 is the sites coupling to cell 2 and so on. The red sites are assigned to the current cell and the lines indicate the sites still to be assigned. The previous cell and all sites already added are indicated by gray and white, respectively. The recursive sweep starting at the final cell and ending in cell 1, indicated by filled arrows, gives the GFs connecting all sites of interest. We can also employ a second recursive sweep, as indicated by the white arrows, to obtain local properties everywhere within the device region.} \label{fig:slice_sketch}
\end{figure*}

To simplify the notation we introduce the dimensionless k-vectors $k_A = 3 k_y a_0 /2$ and $k_Z = \sqrt{3}k_x a_0/2$ such that $f(k_A,k_Z) =1+2\cos\big(k_Z\big)\ee^{\ii k_A} $ , and write the separation vector in terms of the lattice vectors $\rrr=\rrr_j -\rrr_i = m \aaa_1 +n\aaa_2$.  Inserting this into \eqref{g0} gives
\bal
G^{0} (z,\rrr) & = \frac{1}{2\pi^2} \int\mathrm{d}k_A \int\mathrm{d}k_ZN_{ij}(z,k_A,k_Z)\nonumber \\
% \left( \begin{matrix}
%z & tf(k_A,k_Z)\\
%tf^*(k_A,k_Z) & z
%\end{matrix} \right) \nonumber \\
&\times \frac{\ee^{\ii k_A (m+n)+ \ii k_Z(m-n)}}{z^2-t^2\big(1+ 4 \cos^2(k_Z) + 4 \cos(k_A)\cos(k_Z)\big)}  . \eqlab{g0dim}
\eal
\eqref{g0dim} can be solved using a two-dimensional numerical integration, but as we require \eqref{g0dim} for each Green's function element individually, we wish to increase the performance by doing one integration analytically using complex contour techniques.

Following the approach of Ref. \onlinecite{Power2011}, we use $k_A$ as complex variable and consider the poles, $q$, of the denominator
\bal
q &=\cos^{-1}\bigg[\frac{\frac{z^2}{t^2} -1-4\cos^2\big(k_Z\big) }{4\cos\big(k_Z\big)}\bigg]. \eqlab{q_ka}
\eal
The sign of the pole must be selected carefully to ensure that it lies within the integration contour, \ie \; $\mathrm{Im}(q) > 0$, for contours in the positive half plane corresponding to the situation $m+n \geq 0$.
Care must also be taken with the additional phase terms that arise for opposite sublattice GFs.

Using the residue theorem and integrating over a rectangular Brillouin zone, $k_A \in [-\pi;\pi]$ and $k_Z \in [-\pi/2;\pi/2]$, we finally reduce \eqref{g0dim} to
\bal\eqlab{q0_AC}
G^{0}(z,\rrr) &= \frac{\ii}{4\pi t^2} \int_{-\tfrac{\pi}{2}}^{\tfrac{\pi}{2}} \mathrm{d}k_Z \frac{N_{ij}(z,q,k_Z)\ee^{\ii q (m+n)+\ii k_Z  (m-n)} }{\cos\big(k_Z\big)\sin\big( q\big)} ,
%\nonumber \\
%&\times \left( \begin{matrix}
%z & tf(q,k_Z)\\
%tf^*(q,k_Z) & z
%\end{matrix} \right).
\eal
with $q$ given by \eqref{q_ka}.
A similar expression to \eqref{q0_AC} can be derived when using $k_Z$ as first integration variable. \cite{Power2011}
The above derivation is based upon a nearest neighbour model, but can be generalised to also include, for example, second nearest neighbour terms \cite{Lawlor2014arxiv} or uniaxial strains. \cite{Power2012}

We can now use \eqref{q0_AC} to calculate the elements of the required GFs, $\GGn^{(\mathrm{0})}_{B,B}$ and $\GGn^{(\mathrm{0})}_{B,D}$, defined in \secref{SquareSelfEnergy}.
In this way, \eqref{q0_AC} can be used to fill up the elements of the desired matrices one at a time.
Since we need GF matrices of size $N_B \times N_B$ and $N_B \times N_D$, where $N_D$ and $N_B$ are the number of sites at the edge of the device region and in the region B, respectively, it could seem very ineffective to calculate one element at a time.
However, the total number of GF elements to be calculated is greatly reduced by the symmetries of the pristine graphene lattice.
The lattice itself is six-fold symmetric and each of these six identical wedges is in turn mirror symmetric, resulting in a $12$-fold degeneracy of the GFs indexed by site separation vectors.
Additionally, many of the required elements in $\GGn^{(\mathrm{0})}_{B,B}$ and $\GGn^{(\mathrm{0})}_{B,D}$ are identical.
For instance, the onsite and nearest neighbour GF element appear many times, but only need to be calculated once.
Taking the device region in \figref{SSE_sketch} as example we have $N_D =N_B = 20$, yielding 400 individual elements for a brute force calculation.
Instead, using symmetries and duplicates, we only need to calculate 38 and 42 elements when determining $\GGn^{(\mathrm{con})}_{B,B}$ and $\GGn^{(\mathrm{con})}_{B,D}$, respectively.
The reduction becomes more significant for larger systems, as we generally only need to add the GF elements corresponding to the longest couplings.
Consequently, only a small percentage of the GF elements need to be calculated individually and their values for frequently used separations and energies can be stored or reused to enable extremely fast calculation of the required self energies.

\subsection{Adaptive recursion for device region} \seclab{recursive}
In this section we consider the device region where the boundary self energy can be added at the edge.
The full GF of the device region is given by $\GGn_{D} = \big(E\mathbf{1} -\HHH_{D} - \mathbf{\Sigma}_{B}\big)^{-1}$, where we have simplified the notation from \eqref{Gdd}.
From this GF both transport and local properties can be obtained.
However, for most purposes we do not require every element of the Green's function matrix element in the device region, and so to avoid a time consuming full matrix inversion, recursive methods are often applied \cite{Areshkin2010,Lewenkopf2013,Yang2011,Thorgilsson2014,Metalidis2005,Cresti2003,Sajjad2013,CostaGirao2013}.

%The standard recursive approach relies on the division of the Hamiltonian into cells. Each cell only couples to its neighbours. Leads $L$ and $L'$ are coupled to the system by including the self energy terms $\Sigma_L$ and $\Sigma_{L'}$ to the first and last cell, respectively. On matrix form this means a Hamiltonian of the form
%\bal
%\left( \begin{matrix}
%\HHH_{1,1} +\mathbf{\Sigma}_L & \VVV_{1,2} &0& 0&\cdots & 0\\
%\VVV_{2,1} & \ddots &\cdots &\cdots & \cdots &0 \\
% \vdots &\VVV_{n-1,n} &\HHH_{n,n} & \VVV_{i,i+1}&\cdots &\vdots\\
%0 & \cdots &\cdots   & \cdots &\ddots & \VVV_{N-1,N}\\
%0&0& 0  & \cdots & \cdots & \HHH_{N,N}+\mathbf{\Sigma}_{L'}
%\end{matrix} \right) .\eqlab{tridiag}
%\eal
%However, this division means that the standard recursive method is normally limited to two-terminal quasi-one-dimensional problems where the tridiagonal form of the Hamiltonian arises naturally from the geometry. On the other hand, we want a method independent of the shape of the device region and the configuration (and number) of leads.

This section outlines an adaptive recursion method which efficiently includes the boundary self energy as well as an arbitrary device region shape and configuration (and number) of leads.
Alternative approaches have been developed to treat arbitrary shaped regions with multiple leads \cite{Kazymyrenko2008,Wimmer2009,CostaGirao2013}.
These so-called knitting-algorithms add single sites at a time.
They rely on a complicated categorizing of sites into different intermediate updating blocks making the theory and implementation cumbersome.
Hence, we use an approach similar to the ones in Refs. \onlinecite{Areshkin2010,Yang2011,Thorgilsson2014}, and employ an adaptive partitioning of the Hamiltonian matrix in order to bring it into the desired tridiagonal form suitable for recursive methods.

%transmission between two leads, $L$ and $L'$, which is given by the Landauer-B{\"u}tikker formula
%\bal
%\calT_{L,L'}(E) = \mathrm{Tr}\big[ \GGn_{L',L} \mathbf{\Gamma}^L_{L,L} \GGn^\dagger_{L,L'}\mathbf{\Gamma}^{L'}_{L',L'}\big],\eqlab{Landauer}
%\eal
%where $\GGn_{L,L'}=\GGn_{L,L'}(E)$ is the GF submatrix coupling the sites connected to lead $L$ and $L'$. The spectral function is given from the self energies of the lead $\mathbf{\Sigma}^L$ so $\mathbf{\Gamma}^L(E) = \ii\big(\mathbf{\Sigma}^L(E)-\mathbf{\Sigma}^{L}(E){}^\dagger\big)$. From \eqref{Landauer} we notice that we only need the block $\GGn_{L',L}$ to calculate the transmission.

Calculating physical properties generally requires certain GFs connecting a specific set of sites in the device region.
These sites of interest, for example, could be sites where we want to introduce defects, or couple to probes for transport calculations, or measure properties like the local density of states.
We focus first on the general partitioning process, and then demonstrate how it can be quickly modified to account for the edge self-energy terms.
We begin by placing all these sites of interest into recursive cell 1, as shown by the red sites in \figref{slice_sketch}.
We emphasize that the cells in this process are not of a fixed size and may consist of arbitrary sites which are not necessarily connected.
Cell 2 is determined by selecting all the remaining unpartitioned sites which couple directly to sites in cell 1 via a non-zero Hamiltonian matrix element.
In the example in \figref{slice_sketch}, this consists of nearest neighbor sites of those in cell 1, which are not themselves in cell 1.
This process is repeated until all sites in the device region have been allocated a cell, and is demonstrated schematically in the panels of \figref{slice_sketch} where red sites indicate the current cell, and dark gray or white sites indicate sites added to the previous cell, or to earlier cells, respectively.

With the resultant block tridiagonal Hamiltonian, we can now employ the usual recursive algorithm, starting from cell $n=N$, so that the final step yields the required GF sites in cell $n=1$.
These terms can then be used to calculate observable quantities like transmission, LDOS, \etc \; Afterwards a reverse recursive sweep from $n=1$ to $n=N$ can be implemented to efficiently map local quantities like bond currents or LDOS everywhere within the device region\cite{Lewenkopf2013}.
For completeness the full recursive method is summarized in \appref{app_recursive} including the reverse sweep.
We emphasize that the presented method is not unique to graphene systems, but can be employed to arbitrary tight-binding-like models.

\begin{figure}[htb]
%\centering
\includegraphics[width=0.48\textwidth]{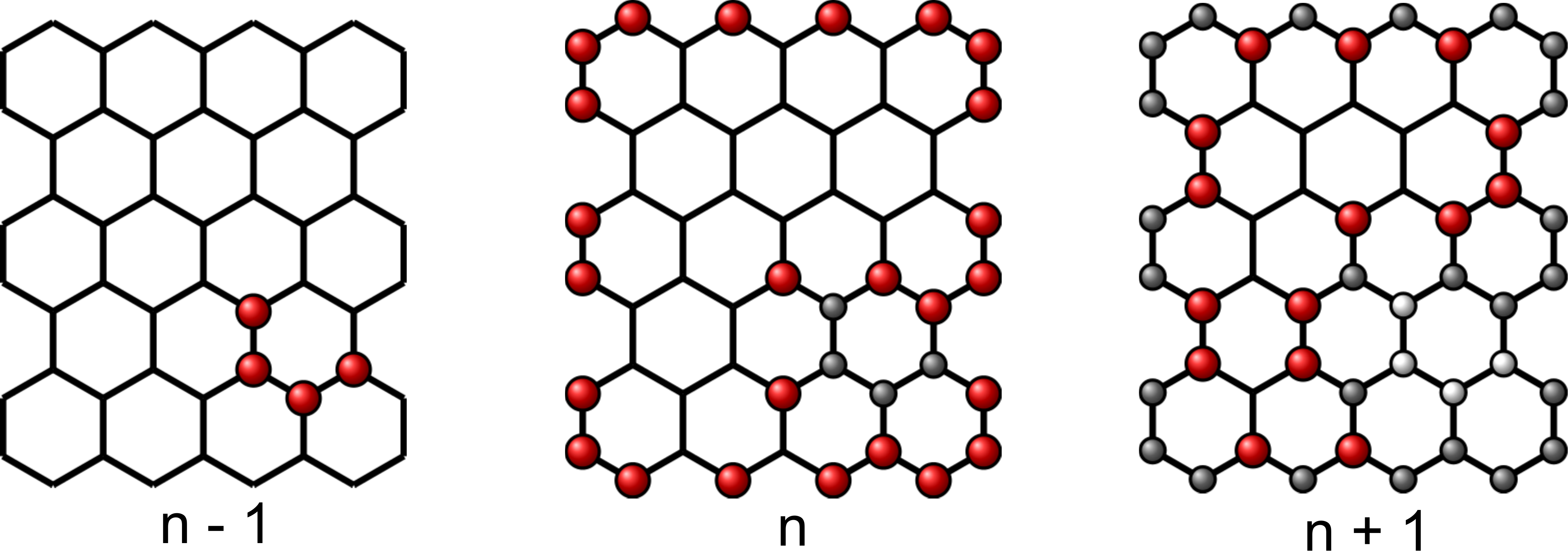}
\caption[]{ An example of the partitioning when the cell $n-1$ is connected to the edge, and we need to include the boundary self-energy, $\mathbf{\Sigma}_{B}$. In this case, all edge sites and self energy terms are included in cell $n$. The symbols are similar to \figref{slice_sketch}.
 } \label{fig:edge_sketch}
\end{figure}

\subsubsection*{Including the boundary self-energy}
We now return to the specific case at hand where the recursive method outlined above needs to be adapted carefully to take account of the boundary self energy.
In general $\mathbf{\Sigma}_{B}$ is a non-hermitian dense matrix connecting all edge sites of the device region.
Therefore it is essential to assign all edge sites to the same cell.
This principle is shown in \figref{edge_sketch}.
If cell $n-1$ contains sites which connect to an edge site, then cell $n$ must contain not only the edge sites directly connecting to cell $n-1$, but also all other edge sites, as these are connected to each other via $\mathbf{\Sigma}_{B}$.
In this way, the cell, $n+1$, must then contain all the sites connecting to cell $n$, \ie \;also connecting to the edge, but not included in cell $n$.
The full cell partitioning algorithm, including this step, is given in \appref{app_recursive}.

\begin{figure}[tb!]
\includegraphics[width=\columnwidth]{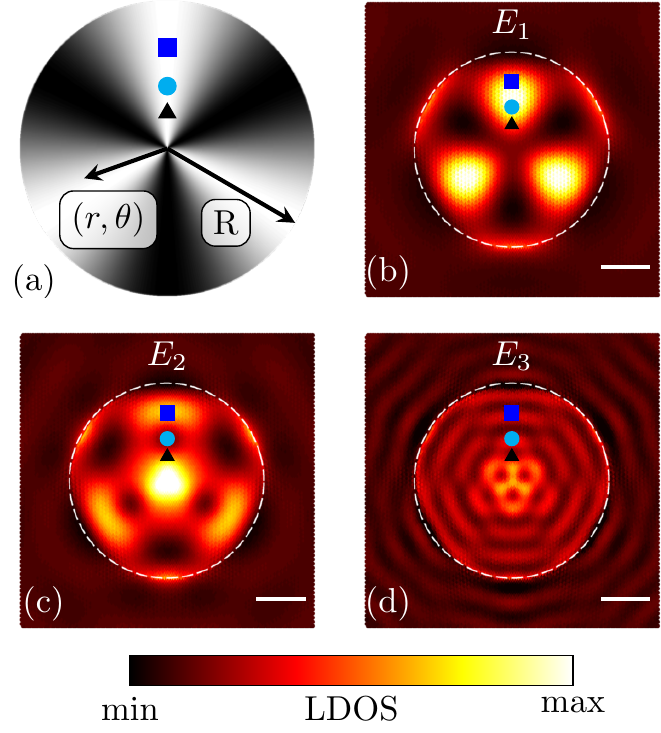}
\caption[]{(a) The PMF distribution calculated using the strain distribution in \eqref{strainprofile}, dark being negative field and light being positive. (b-d) Real space LDOS maps for the A sublattice taken at the energies $E_1=0.06|t|$, $E_2=0.089|t|$ and $E_3 =0.23|t|$, corresponding to energies of the first two pseudo Landau levels and an energy dominated by Friedel type oscillations, respectively. The energies and the symbols correspond the ones used in \figref{ldos_sts}. Sublattice B is similar and is obtained by rotating $60^{\circ}$. The scale bar is 5 nm.
 } \label{fig:PMF}
\end{figure}

\section{Inhomogeneous strain fields in graphene bubbles} \seclab{bubble}
In this section, we employ the patched Green's function method to a locally strained graphene system, demonstrating how it can prove a useful tool in investigating local properties of non-periodic nanostructures in extended two dimensional systems.

Strain engineering has been proposed as a method to manipulate the electronic, optical and magnetic properties of graphene. \cite{Low2011,Guinea2009,Jones2014, Zenan2014,Neek-Amal2012a,Lu2012,Carrillo-Bastos2014,Juan2011,Neek-Amal2013,Neek-Amal2012b,Pereira2009,Pereira2009,Pereira2009prl,Power2012,Pereira2010,Moldovan2013}
It is based on the close relation between the structural and electronic properties of graphene.
The application of strain can lead to effects like bandgap formation \cite{Pellegrino2010}, transport gaps \cite{Low2011} and pseudomagnetic fields (PMFs). \cite{Guinea2009,Zenan2014,Jones2014}

Uniaxial or isotropic strain will not produce PMFs, although it has been shown to shift the Dirac cone of graphene and induce additional features in the Raman signal. \cite{Ni2008}
On the other hand, inhomogeneous strain fields can introduce PMFs.
In this case, the altered tight binding hoppings mimic the role of a gauge field in the low energy effective Dirac model of graphene. \cite{Suzuura2002,Vozmediano2010}
For example Guinea \etal \; \cite{Guinea2009} demonstrated that nearly homogeneous PMFs can be generated by applying triaxial strain.
One of the most striking consequences of homogeneous PMFs is the appearance of a Landau-like quantization. \cite{Guinea2009,Neek-Amal2013}
Scanning tunnelling spectroscopy on bubble-like deformations see this quantization, where the observed pseudo-Landau levels corresponds to PMFs stronger than 300 T. \cite{Lu2012,Levy2010}

Deformations can be induced in graphene samples by different techniques like pressurizing suspended graphene \cite{Bunch2008,Zenan2014} or by exploiting the thermal expansion coefficients of different substrates.\cite{Lu2012}
As a result, introducing nonuniform strain distributions at the nanoscale is a promising way of realizing strain engineering.
The standard theoretical approach to treat strain effects employs continuum mechanics to obtain the strain field.
Several studies improve the accuracy by replacing the continuum mechanics by classical molecular dynamics simulations. \cite{Neek-Amal2012a,Zenan2014,Qi2013}
The strain field can then be coupled to an effective Dirac model of graphene to study the generation of PMFs in various geometries.
In most studies, only the PMF distribution is considered as opposed to experimentally observable quantities like local density of states.
The framework presented in \secref{method} enables us to treat the effect of strain on the LDOS directly from a tight-binding Hamiltonian. Consequently, we are now able to describe a single bubble in an extended system without applying periodic boundary conditions which may introduce interactions between neighboring bubbles.
The dual recursive sweep then allows for efficient calculation of local properties everywhere in the device region surrounding a bubble, enabling us to investigate spatial variations in real space LDOS maps.
In this section we only treat one nanostructure, but the patched Green's function technique efficiently handles several spatially separated nanostructures, as the separation is added very efficiently through the self energy term.

To account for strain within a tight binding approach we modify the hopping parameters.\cite{Pereira2009,Moldovan2013,Carrillo-Bastos2014} The nearest neighbour hopping in \eqref{H0} between site $i$ and $j$ is given by the new distance, $d_{ij}$, between the sites,
\bal
t_{ij} = t \ee^{-\beta \big(\fracsmall{d_{ij}}{a_0} -1\big)},
\eal
where the coefficient $\beta = -\partial \;\mathrm{ln}\; t / \partial\; \mathrm{ln}\; a_0 \approx 3.37$. \cite{Pereira2009}
We treat the deformation problem by applying an analytical displacement profile $(u(x,y),z(x,y))$ matched against experimental data for pressurized suspended graphene. \cite{Yue2012}
Here $u(x,y)$ and $z(x,y)$ are the in-plane and vertical displacements, respectively, which are induced by the applied strain.
For a rotationally symmetric aperture with radius $R$, these are given, in spherical coordinates $(r,\theta)$, as
\bsub \eqlab{strainprofile}
\bal
z(r,\theta) & = h_0 \bigg(1-\frac{r^2}{R^2}\bigg),\\
u(r,\theta)& = u_0\frac{r}{R}\bigg(1-\frac{r}{R}\bigg),
\eal
\esub
for $r<R$.
Here $h_0$ is the maximal height of the bubble and $u_0 = 1.136 h_0^2/R$ is a constant relating the out-of-plane and in-plane deformations. \cite{Yue2012}
We note that this profile gives rise to a sharp edge at $r=R$, and many of the features we discuss below emerge from the strongly clamped nature of this bubble type.

As shown in \appref{PMF}, rotationally symmetric strain profiles give rise to threefold symmetric PMFs in the effective Dirac model. This is shown in \figref{PMF}a for the strain profile considered in \eqref{strainprofile}. As discussed in earlier studies, \cite{Moldovan2013,Carrillo-Bastos2014} we get an asymmetric sublattice occupancy such that the LDOS of each sublattice has a threefold symmetric distribution following the PMF while rotated $60^\circ$ compared to the opposite sublattice. In all calculations below, we therefore show only one sublattice, as the result for the opposite sublattice can be obtained by a $60^{\circ}$ rotation and the total pattern is a superposition of both. \cite{Jones2014,Juan2011}

Comparing the PMF distribution in \figref{PMF}a with the calculated LDOS maps at different energies in \figref{PMF}b-d for a bubble of radius $R=10$ nm and height $h_0=3$ nm, we immediately notice that the threefold symmetry is also present in the LDOS maps.
However, the spatial LDOS maps have significant additional details compared to the PMF distribution.

\begin{figure}[tb]
\includegraphics[width=0.4\textwidth]{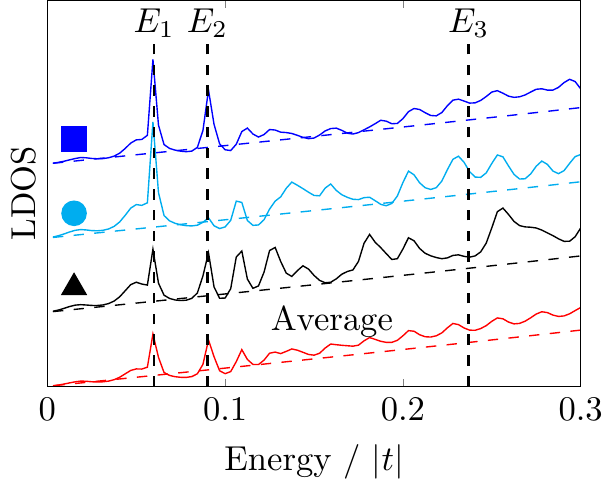}
\caption[]{ The LDOS as a function of energy for the three positions indicated in \figref{PMF} and for the average of the `slice' of the bubble region containing the symbols. The dashed lines indicate the LDOS without the bubble. The curves are shifted with respect to each other to increase visibility.
 } \label{fig:ldos_sts}
\end{figure}

\begin{figure}[tb]
\includegraphics[width=0.4\textwidth]{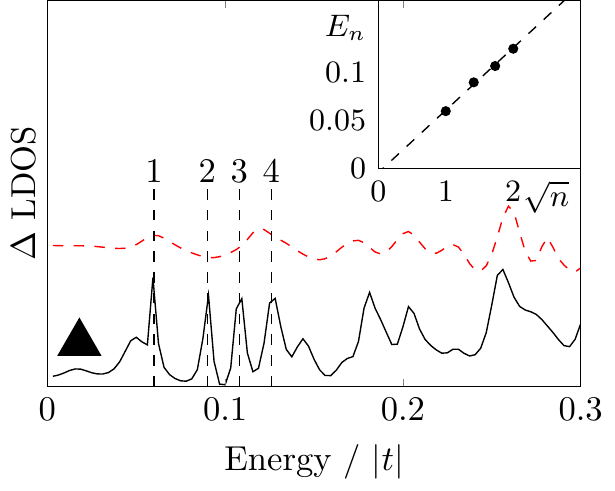}
\caption[]{ The difference in LDOS as a function of energy for the point indicated with a triangle on \figref{PMF}. We show both the full calculation (full line) and an artificial system containing only the perturbation for a small region at the edge of the bubble (dashed line). We adjust the average hopping constant in the calculation of the artificial system to match the full calculation. Inset: The peak energies 1-4 as a function of $\sqrt{n}$, where $n$ is the peak number.
 } \label{fig:ldos_sts_test}
\end{figure}

In \figref{ldos_sts} we calculate the energy dependent LDOS at the positions indicated by symbols (square, circle and triangle) in \figref{PMF}.
We first consider the average of the LDOS within the `slice' containing the symbols, shown by the bottom (red) curve in \figref{ldos_sts}.
Two distinct oscillation types are observed, and we argue that these can be divided into Friedel-type and PMF-induced features.
At high energies in particular we notice regularly space oscillations with an approximate period of  $\hbar v_F \pi/2R$.
These are consistent with Friedel-type oscillations related to the size of the structure and emerging from interferences between electrons scattered at opposite sides of the bubble.
An exact treatment needs to take into account the renormalized Fermi velocity, $v_F$, due to the average change in bond length. \cite{Pellegrino2011}
At lower energies we observe distinct peaks which are not equally spaced (the first two appear at $E_1$ and $E_2$).
We will show that these are due to pseudomagnetic effects and we refer to them as pseudo Landau levels.

Besides the Friedel oscillation associated with the bubble radius, we also have similar oscillations associated with the distances to different edges of the bubble.
These features are highly position dependent, and explain the differences between the three single position curves in \figref{ldos_sts}.
When considering the average, these position dependent oscillations are washed out (bottom curve in \figref{ldos_sts}), leaving only the oscillation dependent on the structure size.
However, at individual positions these oscillations can have a considerable impact.
Returning to the individual position STS curves in \figref{ldos_sts}, we note that the peak at $E_2$ is only dominant for the points indicated by the square and triangle.
It is suppressed by Friedel-type interferences at the circle point, which is also clear from the LDOS map in \figref{PMF}c.

The amplitude of the Friedel-type oscillations is determined by the strength of scattering near the bubble edges.
The clamped edge implied by the strength profile in \eqref{strainprofile} gives rise to significant strain fields along this edge, leading to a sharp, strong perturbation.
More realistic profiles calculated from molecular dynamics also indicate strong perturbations near the edges of clamped bubbles.\cite{Zenan2014}
Our results indicate that edge scattering effects may significantly affect LDOS behavior in clamped bubble systems and even mask PMF-induced features.

To treat the oscillations due to the feature size and edge sharpness in more detail, we calculate the LDOS for an artificial system only taking into account the strain field along a small ring around the edge, see \figref{ldos_sts_test} (dashed red line).
In this way, only Friedel-type features are expected within the structure.
If we compare to the full calculation (full black line in \figref{ldos_sts_test}), we notice that the oscillations at higher energies are present in both calculations, whereas the sharp peaks are only present in the full calculation.
This confirms the Friedel nature of the higher energy oscillations and suggests the lower energy peaks are due to an alternative mechanism.
To confirm that the sharp peaks are due to pseudomagnetic effects, we compare the peak positions to the standard form expected for Landau levels in graphene $E_n = \mathrm{sign}(n)\sqrt{2 e_0 \hbar v_F^2 B_s n} $, where $e_0$ is the electron charge, $B_s$ is the magnetic field and $n$ is the peak number. \cite{CastroNeto2009}
The peaks labelled 1-4 in \figref{ldos_sts_test} display the $\sqrt{n}$ dependence characteristic of Landau levels in graphene, as shown in the inset of \figref{ldos_sts_test}. The size of the PMF can furthermore be inferred to be $B_s\sim 30$ T from the inset.

To conclude, we discussed how the features in the LDOS spectra of clamped graphene bubbles can be explained by a combination of size-dependent scattering and PMF-induced effects like pseudo Landau quantization.
Significant strain fields near the edge of the structure give rise to strong Friedel-type oscillations in the LDOS and these oscillations envelope the effect of a PMF.
We must therefore be careful to distinguish between the two type of oscillations when investigating the electronic effects of PMFs induced by inhomogeneous strain fields.

\begin{figure}[tb]
\centering
\includegraphics[width=0.85\columnwidth]{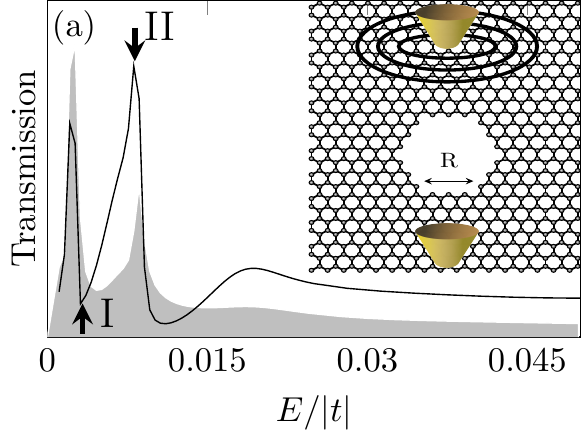}
\includegraphics[width=1\columnwidth]{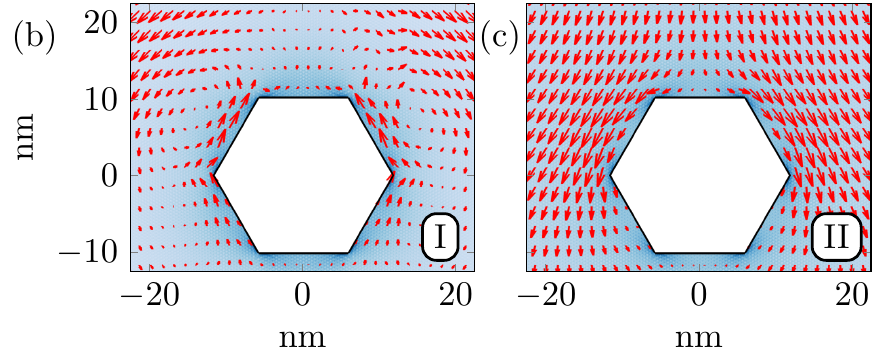}
\caption[]{\small (a)  The transmission as a function of energy for a dual probe setup with an antidot in between the probes as schematically shown in the inset. The distance between the probes are 200 nm and the antidot with purely zigzag edges has side length $R=48a_0\sim 6.8$ nm. The shaded area corresponds to the LDOS around the edge of the antidot.
antidot } \label{fig:Antidot_trans}
\end{figure}

\section{Vortex currents near perforations} \seclab{antidot}
In this section we investigate local transport properties near antidots (\ie \;perforations) in a graphene sheet.
Periodic arrays of antidots have been studied as a way to open a bandgap in graphene \cite{Pedersen2008,Gunst2011,Furst2009} or to obtain waveguiding effects. \cite{Pedersen2012,Power2014}
Furthermore, a single perforation in a graphene sheet has been considered as a nanopore for DNA sensing. \cite{Schneider2010,Merchant2010}

Several studies show that the electronic structure of antidots is closely related to the exact edge geometry. \cite{Power2014,Pedersen2008,Settnes2014b}
Experimental fabrication techniques like block copolymer \cite{Kim2010,Bai2010,Kim2012} or electron beam lithography, \cite{Eroms2009,Qiang2013,Oberhuber2013} inevitably lead to disorder and imperfect edges.
However, it may be possible to control the edge geometry of the antidot by heat treatment \cite{Jia2009,Qiang2013}, or selective etching. \cite{Oberhuber2013,Pizzocchero2014}

Motivated by the interest in how current flows in antidot systems, we apply the patched GF method to a single perforation in a graphene sheet. The method allows us to study the perforation with no influence from periodic repetition or finite sample size.
Additionally, the combination of recursive methods and a boundary self-energy allows for investigation of antidot sizes realizable experimentally. \cite{Cagliani2014,Bai2010,Merchant2010}
In fact we consider both an example antidot with perfect edges and an exact structure found from high resolution transmission electron microscope (TEM) images using pattern recognition. \cite{Kling2014,Vestergaard2014}

To investigate current on the nanoscale, recent experiments have realized multiple STM-systems. \cite{Baringhaus2014,Sutter2008,Clark2014,Li2013}
These allow for individual manipulation of several STM-tips in order to make electrical contact to the sample near the considered nanostructure.
Theoretically, we previously considered multiple STM setups allowing for both fixed and scanning probes. \cite{Settnes2014a,Settnes2014b}
The method presented here allows for not only transmission calculations but also calculation of local electronic and transport properties in the presence of multiple point probes.
At the same time large separations between the different probes and/or nanostructures are easily included as additional separation is achieved in a very computationally efficient manner through the self energy term connecting multiple patches.
The combination of large spatial separation between features, while still enabling calculation of local electronic and transport properties, can prove a useful tool in investigating extended two dimensional systems where we take special interest in a particular region of the extended sample.

In order to consider transmissions and current patterns, we add leads to the system through inclusion of a lead self-energy term, $\Sigma^{L}_{ij} = V^L_{is} g^s(\rrr_i-\rrr_j) V^L_{sj}$, where $V^L_{is}$ is the coupling element between the device site $i$ and the lead.
To model the structureless lead, we use the surface GF of a single atomic chain, as this has a constant DOS in the considered energy range. The distance dependence in $g^s(\rrr_i-\rrr_j)$ is necessary to avoid an unphysical coupling between different lattice sites via the lead. We therefore add a $1/|\rrr_i-\rrr_j|$-dependence for the off-diagonal terms \footnote{The surface Green's function of the single atomic chain is given by  $g^s = \frac{E\pm\sqrt{E^2-4\gamma^2}}{2\gamma^2}$. \cite{ecoBook} We chose $\gamma=|t|$ as this gives a constant DOS within the considered energies. The distance dependence for off-diagonal finally gives the GF, $g^s(\rrr_i-\rrr_j) = \delta_{ij}g^s +(1-\delta_{ij})\frac{g^s}{|\rrr_i-\rrr_j|}$ } where $\rrr_i\neq \rrr_j$, as appropriate for a structureless three-dimensional free electron gas. \cite{ecoBook}

\begin{figure}[tb]
\centering
\includegraphics[width=0.95\columnwidth]{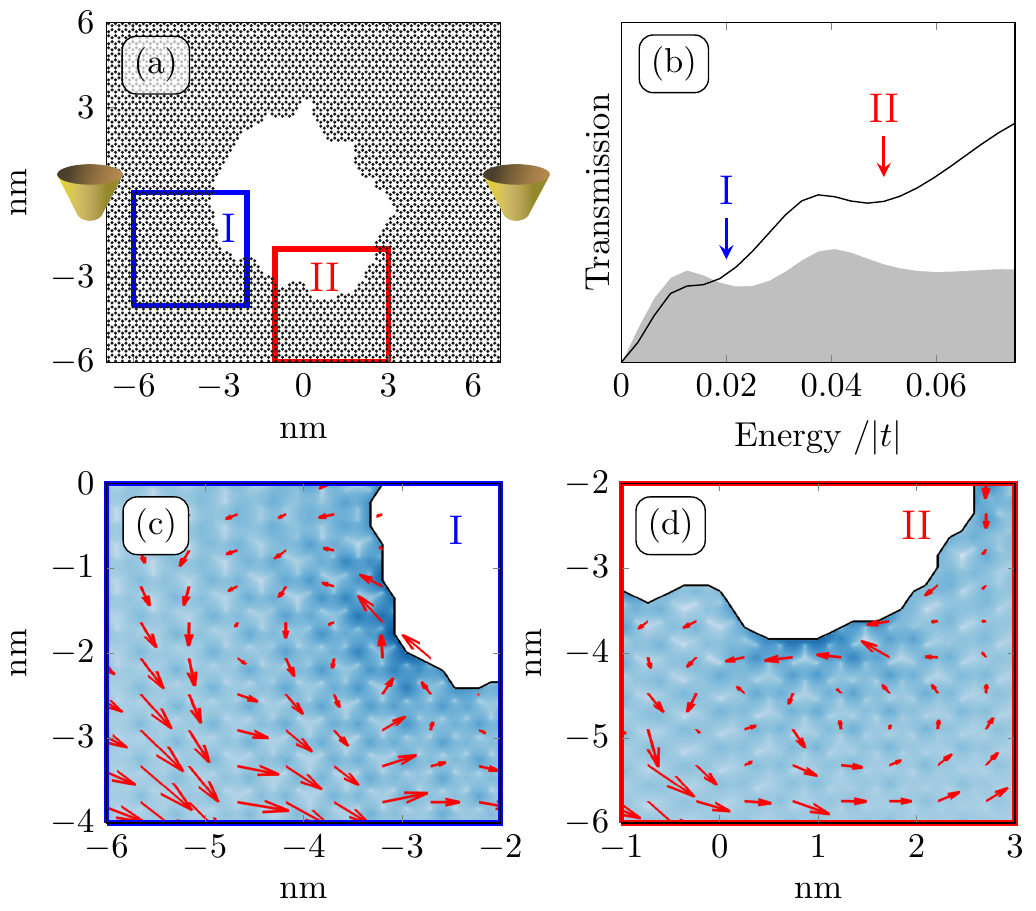}
\caption[]{\small An actual perforation is obtained from high resolution TEM images through pattern recognition and we consider the vortex like current paths forming around the perforation as certain energies. (a) Shows the structure of the perforation as well as an indication of the probe position (in the actual calculations the probes are 200 nm apart). The indicated areas corresponds to the zooms in (c) and (d). (b) The transmission for the dual probe setup. The shaded area indicate the average LDOS around the edge of the antidot. Furthermore, the energies I and II corresponding to the energies used on (c) and (d), respectively. (c-d) Bond current maps taken at the energies I and II, respectively, and shown at the positions indicated on (a).
 } \label{fig:TEM}
\end{figure}

First, we consider a zigzag-edged antidot with side length $R=48 a\sim 12$ nm, where $a$ is the length of the graphene unit cell and $a = \sqrt{3} a_0=2.46$ $\textrm{\AA}$. This is comparable to experimental sizes where sub-20-nm feature sizes have been reported. \cite{Merchant2010,Kim2010,Bai2010,Kim2012}
The antidot is between two probes placed $200$ nm apart, as shown schematically in the inset of \figref{Antidot_trans}a.
The main panel of \figref{Antidot_trans}a shows the transmission as a function of energy for this dual point probe setup.
We note the distinct transmission peaks.
As explained in Ref. \onlinecite{Settnes2014b}, these peaks are related to localized states along the zigzag edges. As a consequence, we notice the correspondence between the peaks in the transmission and the peaks in the LDOS around the edge, see shaded area in \figref{Antidot_trans}a.

Next, we calculate the bond currents from the top lead.
The bond current between site $i$ and $j$ from lead $L$ are calculated, as explained in \appref{app_recursive}, by $J^L_{ij} = -H_{ij}\mathrm{Im}\big[\GGn^a\mathbf{\Gamma}^L\GGn^r\big]_{ij}/\hbar$, where $H_{ij}$ is the Hamiltonian matrix element connecting site $i$ and $j$.
The bond currents around the zigzag antidot for the energies indicated in \figref{Antidot_trans}a are shown in Figs. \ref{fig:Antidot_trans}b and \ref{fig:Antidot_trans}c.
In this way, we see that the transmission dips are related to vortex like current paths.
These vortex paths create a larger  `effective size' for the antidot at this energy, characterized by a region around the antidot avoided by the current paths.
On the other hand, at the transmission peaks the current passes near to the antidot edge.

The antidot considered in \figref{Antidot_trans}, although of realistic size, is an idealization, as experimental perforations will inevitably contain imperfections.
To consider a more realistic case, we turn to a perforation observed in experimental TEM images.
Using pattern recognition \cite{Kling2014,Vestergaard2014} the positions of the individual carbon atoms are obtained from high resolution TEM images (see Fig. 9 of Ref. \onlinecite{Vestergaard2014}).
Pristine graphene is added around the experimentally obtained perforation to obtain the system shown in \figref{TEM}a.
From the transmission (see \figref{TEM}b), we notice that peaks are still present, but broadened by the disorder.
Considering the two energies I and II in \figref{TEM}b and comparing their spatial current maps, we find that certain positions around the antidot are responsible for the additional backscattering causing the transmission dips.
Dip I corresponds to a vortex pattern at the left side of the antidot (see \figref{TEM}c), whereas the dip at II is caused by a vortex pattern at the bottom of the antidot (see \figref{TEM}d).
This result suggests that electrons at different energies see a different effective perforation size and shape and are scattered accordingly.

\section{Conclusion}
We have expanded the standard recursive Green's function method to calculate local and transport properties enabling calculations in extended non-periodic systems.
We exploit an efficient calculation of the pristine two-dimensional GF using complex contour methods.
Once calculated, the pristine GFs are used to determine a boundary self energy term describing the extended system.
In this way, we can treat a finite device region embedded within an extended sample.

We first demonstrated how this approach is able to efficiently treat the electronic properties of strained bubbles in an extended graphene sheet.
Considering a clamped bubble, we have shown that the finite size gives rise to Friedel-type oscillations in the density of states.
This effect mixes with any pseudomagnetic effects arising from the strain field. We show that the edge effects can cloud pseudomagnetic signatures in the LDOS by adding additional structure which is not directly related to pseudomagnetic effects.

Secondly, we showed how finite leads can be added to a patched device region to efficiently calculate transport properties for spatially separated features, while still being able to map local properties in various parts of the system.
In particular, we investigated the current flow around perforations of a graphene lattice.
Both idealized geometries and experimental geometries obtained from high resolution TEM images were considered.
The transmissions show distinct dips caused by localized states along zigzag segments of the perforations. The transmission dips were associated with vortex-like current paths formed near the perforation edges.

We have demonstrated the versatility of this novel approach to the popular recursive GF method.
The method allows for calculation of the same local and transport properties as standard methods, but adds the ability to treat large non-periodic structures embedded in extended samples. We can extend the present method beyond nearest neighbor and to relevant alloys like hBN or transition metal dichalcogenides. We therefore predict that the patched Green's function method will prove a valuable tool in the investigation of nanostructures in two dimensional materials.

\textbf{Acknowledgements}
We thank B.K. Nikolic for enlightening discussions of recursive methods applied to arbitrary geometries. We also thank J. Kling for providing the experimental TEM data used in modelling current paths around realistic perforations. The work was supported by the Villum Foundation, Project No. VKR023117.
The Center for Nanostructured Graphene (CNG) is sponsored by the Danish Research Foundation, Project DNRF58.

\appendix

\begin{figure*}[htb]
%\centering
\includegraphics[width=0.8\textwidth]{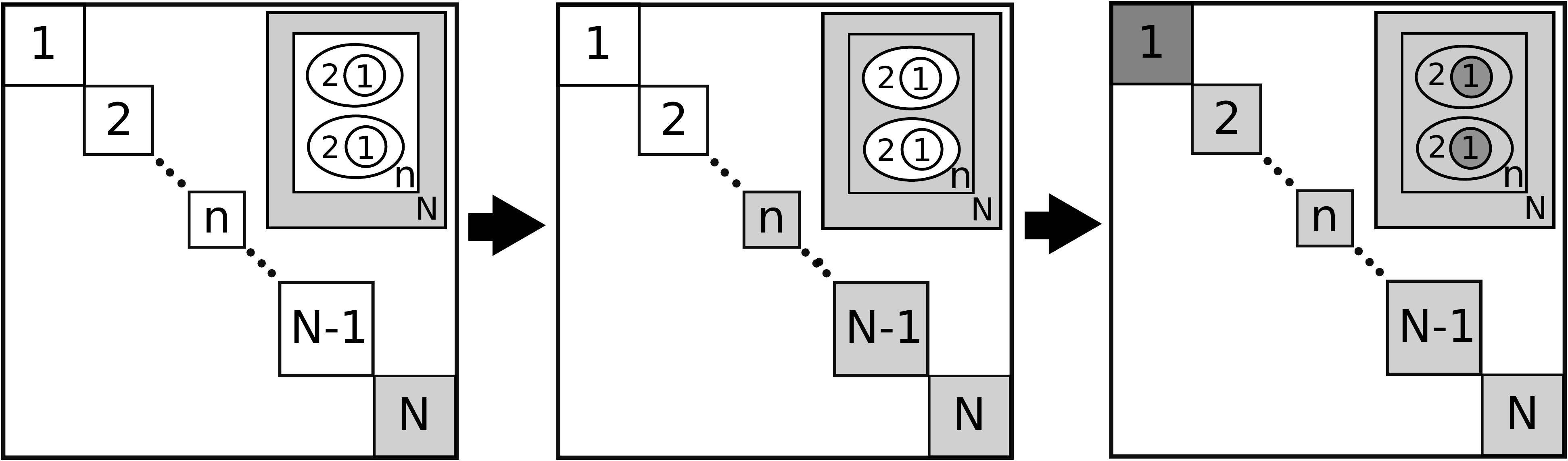}
\includegraphics[width=0.8\textwidth]{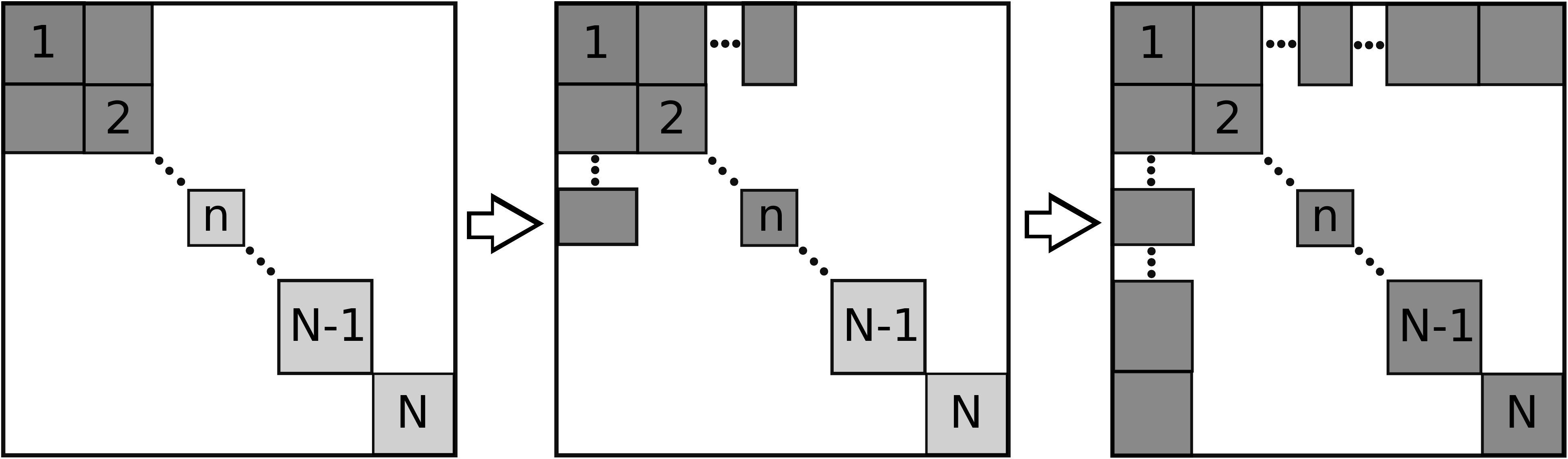}
\caption[]{Top row: recursive sweep going from cell $n=N$ to $n=1$. Light gray indicate blocks that are stored for the reversed sweep and dark gray indicate blocks of the full GF. The inset shows an illustration of how the different blocks correspond to neighboring cells in the device region. Bottom row: reversed recursive sweep going from $n=1$ 1 to $n=N$ showing how this sweep can obtain both diagonal and off-diagonal blocks.
 } \label{fig:sweep1}
\end{figure*}

\section{Recursive Algorithm}\seclab{app_recursive}
To obtain a tridiagonal Hamiltonian we let cell $n=1$ contain all sites of interest. Then following the algorithm outlined below we assign all sites into cells.

\begin{itemize}
\item[1:]  Let $\{n\}$ denote all sites in cell $n$ and $\{"unassigned"\}$ denote all sites not yet assigned to a cell.
\item[2:] Find all sites $j$ for which $H_{nj} \neq 0$ where $n\in \{n\}$ and $j\in \{"unassigned"\}$. Denote these sites $\{n+1\}$.
\item[2a:] If $\{n+1\}$ contains an edge site, then all remaining edge sites are added to $\{n+1\}$.
\item[3:] Sites in $\{n+1\}$ are removed from $\{"unassigned"\}$
\item[4:] Repeat 1-3 until all sites are assigned to a cell.
\end{itemize}
Step 2a is included if we require an edge self energy term $\Sigma_B$ as described in \secref{method}.

Assuming the block tridiagonal partitioning obtained from the algorithm above, we make an update sweep starting from cell $n=N$, as shown schematically in \figref{sweep1}. The steps are calculated using the recursive relations \cite{Lewenkopf2013}
\bsub
\bal
\ggn_{N,N} &= \big(E - \HHH_{N,N}\big)^{-1},\\
\ggn_{n,n}& = \big(E - \HHH_{n,n}-\VVV_{n,n+1}\ggn_{n+1,n+1}\VVV_{n+1,n}\big)^{-1},\\
\ggn_{1,1} & = \big(E - \HHH_{1,1}-\VVV_{1,2}\ggn_{2,2}\VVV_{2,1}-\sum_{m=1}^M \Sigma^m_{lead}\big)^{-1},
\eal
\esub
where one of the $\HHH_{n,n}$ terms includes the self energy and $\Sigma^m_{lead}$ terms are included if we calculate transmission.
After the sweep is complete, the fully connected GF of cell $n=1$ is obtained as $\GGn_{1,1} = \ggn_{1,1}$. As all sites of interest are placed in this cell, we can now calculate observables involving these sites. For example we calculate transmission, $\calT_{L,L'}$, between lead L and L' using these GFs.
\bal
\calT_{L,L'}(E) = \mathrm{Tr}\big[ \GGn_{L',L} \mathbf{\Gamma}^L_{L,L} \GGn^\dagger_{L,L'}\mathbf{\Gamma}^{L'}_{L',L'}\big],\eqlab{Landauer}
\eal
where $\Gamma^L = \ii (\Sigma^L-\Sigma^L{}^\dagger)$ and $\GGn_{L,L'}$ ($\GGn_{L,L'}^\dagger$) is the retarded (advanced) GF connecting the two leads L and L'.
% It is noted that the choice of sites in cell $n=1$ is flexible and can be modified to a variety of calculation purposes \ie \; not only transmission calculations.
% For example if we only want the local density of states for a few sites or the correlation between two sites in the device, these sites can easily be included in cell $i=1$.

In order to obtain other blocks of the full GF matrix, we need to store the GF matrix, $\ggn_{n,n}$, for each cell as we sweep from $n=N$ to $n=1$. The stored blocks are shown in light gray on \figref{sweep1}.

To obtain the LDOS at site $i$, $\rho_{ii} = - \mathrm{Im}\big(G_{ii}\big)/\pi$, we need the diagonal of the GF matrix. We calculate the block diagonal from a reversed sweep from $n=1$ to $n=N$, see \figref{sweep1}. The reversed sweep uses the block diagonals, $\ggn_{n,n}$, from the first sweep to calculate the full diagonal GF, $\GGn$,
\bal
\GGn_{n,n}& = \ggn_{n,n} + \ggn_{n,n}\VVV_{n,n-1}\GGn_{n-1,n-1}\VVV_{n-1,n}\ggn_{n,n}.
\eal

Finally, we want to obtain bond currents for the state leaving a lead $L$. This can be calculated by $J^L_{ij} = -H_{ij}\mathrm{Im}\big[\GGn_{i,1}\mathbf{\Gamma}^L_{1,1}\GGn^\dagger_{1,j}\big]/\hbar$. Remembering that the leads are assigned to cell $n=1$, we need the off-diagonal blocks, $\GGn_{1,n}$ and $\GGn_{n,1}$, in order to obtain bond currents. Using the stored GFs from the first sweep we can calculate the needed off-diagonals,
\bsub
\bal
\GGn_{1,n} &= \GGn_{1,n-1}\VVV_{n-1,n}\ggn_{n,n},\\
\GGn_{n,1} &= \ggn_{n,n}\VVV_{n,n-1}\GGn_{n-1,n}.
\eal
\esub

\section{Pseudomagnetic field for rotational symmetric strain field}\seclab{PMF}
The strain tensor is generally given as
\bal
\eps_{ij} = \frac{1}{2} \bigg(\partial_j u_i +\partial_i u_j + (\partial_i z)(\partial_j z)\bigg), \quad i,j = x,y,
\eal
where $\uuu(x,y)$ is the in-plane deformation field and $z(x,y)$ is the out-of-plane deformation.\cite{Vozmediano2010}

A general two dimensional strain field, $\eps_{ij}(x,y)$, leads to a gauge field in the effective Dirac Hamiltonian of graphene \cite{Suzuura2002,Vozmediano2010}
\bal \eqlab{Afield}
\AAA =-\frac{\hbar\beta}{2e a_0} \left( \begin{matrix}
\eps_{xx}-\eps_{yy} \\
-2\eps_{xy}
\end{matrix} \right),
\eal
which in turn gives a PMF
\bal \eqlab{Bs}
B_s &= \rot \AAA = \partial_x A_y - \partial_y A_x. %\nonumber \\
%&=\frac{1}{r} \partial_\theta A_r - \partial_r A_\theta - \frac{A_\theta}{r}.
\eal
\eqsref{Afield}{Bs} imply that the x-axis is chosen along the zigzag direction of the graphene lattice.

Now restricting ourselves to rotationally symmetric deformations, $u(r)=u_r$ and $z(r)=z$, while using polar coordinates $(r,\theta)$ yields
\bal
B_s = -\frac{\hbar\beta}{2ea_0}\bigg(2\frac{g(r)}{r} - \partial_r g(r) \bigg) \sin(3\theta), \eqlab{Bs_rot}
\eal
with $g(r) = \partial_r u_r -u_r/r+\fracsmall{1}{2} \big(\partial_r z\big)^2$.
We notice from \eqref{Bs_rot} that the PMF for a rotationally symmetric displacement field  is always 6-fold symmetric. On the other hand, the magnitude depends on both the in-plane and out-of-plane displacement.

Considering the displacement field in \eqref{strainprofile} we now obtain a PMF of the form,
\bal
B_s = -\frac{\hbar\beta u_0}{2ea_0R^2} \sin(3\theta). \eqlab{Bs_membrane}
\eal
Taking into account the scaling $u_0 \propto h_0^2/R$, we obtain a final scaling of the PMF with the size of the bubble, $B_s \propto h_0^2/R^3$.

%Start from the last layer and end in the first where the leads are placed
%
%Transmission from $\GGn_{1,1} = \ggn_{1,1}$
%\bal
%\calT_{L,L'}(E) = \mathrm{Tr}\big[ \GGn(E) \mathbf{\Gamma}^L(E) \GGn^\dagger(E)\mathbf{\Gamma}^{L'}(E)\big]
%\eal
%\bal
%\mathbf{\Gamma}^L = \ii\big(\mathbf{\Sigma}^L-\mathbf{\Sigma}^{L}{}^\dagger\big)
%\eal
%Block diagonals of Green's function (LDOS)
%\bal
%\GGn_{n,n}& = \ggn_{n,n} + \ggn_{n,n}\VVV_{n,n-1}\GGn_{n-1,n-1}\VVV_{n-1,n}\ggn_{n,n}
%\eal
%Top part of Green's function matrix $\GGn_{1,i}/\GGn_{i,1}$
%\bsub
%\bal
%\GGn_{1,n} &= \GGn_{1,n-1}\VVV_{n-1,n}\ggn_{n,n}\\
%\GGn_{n,1} &= \ggn_{n,n}\VVV_{n,n-1}\GGn_{n-1,n}
%\eal
%\esub
%
%Bond Currents of the form $J^L_{ij} = -H_{ij}\mathrm{Im}\big(\GGn\mathbf{\Gamma}^L\GGn^\dagger\big)_{ij}/\hbar$. Define $\calG = \GGn\mathbf{\Gamma}^L\GGn^\dagger$. For in-cell currents from lead L
%\bsub
%\bal
%\calG^{L}_{1,1} &= \GGn_{1,1}\mathbf{\Gamma}^{L}_{1,1}\GGn^\dagger_{1,1}\\
%\calG^{L}_{n,n} &= \ggn_{n,n}\VVV_{n,n-1}\calG^{L}_{n-1,i-1}\VVV_{n-1,n}\ggn_{n,n}^\dagger
%\eal
%\esub
%Inter cell current obtained by
%\bal
%\calG^{L}_{n,n-1} & = \ggn_{n,n}\VVV_{n,n-1}\calG^{L}_{n-1,n-1}
%\eal

%\bibliography{Bibfile2}

\end{document}